\documentclass[journal,compsoc]{IEEEtran}
\usepackage{graphicx,times,amsmath,amssymb,cite,subfigure,stfloats,booktabs, url,multirow,afterpage}
\usepackage[lined,ruled]{algorithm2e}
\allowdisplaybreaks

\makeatletter

\newcommand{\Rmnum}[1]{\expandafter\@slowromancap\romannumeral #1@}
\makeatother

\begin{document}
\title{CMCRD: Cross-Modal Contrastive Representation Distillation for Emotion Recognition}

\author{
    Siyuan Kan, Huanyu Wu, Zhenyao Cui, Fan Huang, Xiaolong Xu and Dongrui Wu,~\IEEEmembership{Fellow,~IEEE}
\IEEEcompsocitemizethanks{
\IEEEcompsocthanksitem S. Kan, H. Wu, Z. Cui and D. Wu are with the Key Laboratory of Image Processing and Intelligent Control, Ministry of Education, School of Artificial Intelligence and Automation, Huazhong University of Science and Technology, Wuhan, 430074, China.
\IEEEcompsocthanksitem F. Huang and X. Xu are with Wuhan Institute of Digital Engineering, Wuhan, 430074, China. F. Huang is also with Shanghai Jiao Tong University, Shanghai, 200240, China.
\IEEEcompsocthanksitem X. Xu and D. Wu are the corresponding authors. Emails: xuxl@cssc-lincom.cn, drwu@hust.edu.cn.}
}

\IEEEtitleabstractindextext{
\begin{abstract}
Emotion recognition is an important component of affective computing, and also human-machine interaction. Unimodal emotion recognition is convenient, but the accuracy may not be high enough; on the contrary, multi-modal emotion recognition may be more accurate, but it also increases the complexity and cost of the data collection system. This paper considers cross-modal emotion recognition, i.e., using both electroencephalography (EEG) and eye movement in training, but only EEG or eye movement in test. We propose cross-modal contrastive representation distillation (CMCRD), which uses a pre-trained eye movement classification model to assist the training of an EEG classification model, improving feature extraction from EEG signals, or vice versa. During test, only EEG signals (or eye movement signals) are acquired, eliminating the need for multi-modal data. CMCRD not only improves the emotion recognition accuracy, but also makes the system more simplified and practical. Experiments using three different neural network architectures on three multi-modal emotion recognition datasets demonstrated the effectiveness of CMCRD. Compared with the EEG-only model, it improved the average classification accuracy by about 6.2\%.
\end{abstract}

}
\maketitle

\section{Introduction}

Emotion, as a crucial medium for human communication with the external world~\cite{barrett2017emotions}, has a huge impact on an individual's perception, communication, decision making, etc. Emotion recognition uses data from one or more modalities to help machine identify and better understand human emotions, which has become an important technology in human-computer interaction and attracted increasing attention~\cite{cowie2001emotion,lin2010eeg,zheng2018emotionmeter,drwuCTIAL2024}.

In emotion recognition, typical modalities used include facial expressions~\cite{zhong2014learning,lee2019context}, speech~\cite{el2011survey,drwuTAC2022}, eye movements~\cite{soleymani2011multimodal}, electroencephalography (EEG)~\cite{zheng2017identifying}, and so on. Among them, EEG provides a direct measure of brain activities and the origin of emotions~\cite{alarcao2017emotions,wu2023affective}, and is more difficult to fake. With the development of deep learning, EEG-based emotion recognition has advanced from traditional machine learning approaches like Support Vector Machine and $k$-Nearest Neighbors~\cite{wang2014emotional} to deep learning approaches, including Convolutional Neural Networks (CNNs)~\cite{zhang2018cascade,lawhern2018eegnet}, Recurrent Neural Networks~\cite{zhang2018spatial,li2020novel,chowdary2022emotion}, Graph Neural Networks (GNNs)~\cite{song2018eeg}, Dynamical Graph Convolutional Neural Networks (DGCNN)~\cite{zhong2020eeg}, multi-head attention mechanisms~\cite{song2022eeg,ding2024eeg}, and so on.

Although numerous EEG-based emotion recognition models have been proposed, the performance of EEG-only models remains unreliable, because emotions are influenced by both internal and external activities of the individual. Therefore, researchers started to consider multi-modal fusion, i.e., combine EEG with other modalities~\cite{zhu2022mutual,fu2023novel}. One example is to combine EEG signals reflecting internal cognitive state with eye movement signals reflecting external subconscious behaviors, which not only improves the accuracy of emotion recognition, but also has high interpretability~\cite{soleymani2011multimodal,lu2015combining,zhu2022content}.

\begin{figure}[htbp]         \centering
\includegraphics[width=.9\linewidth,clip]{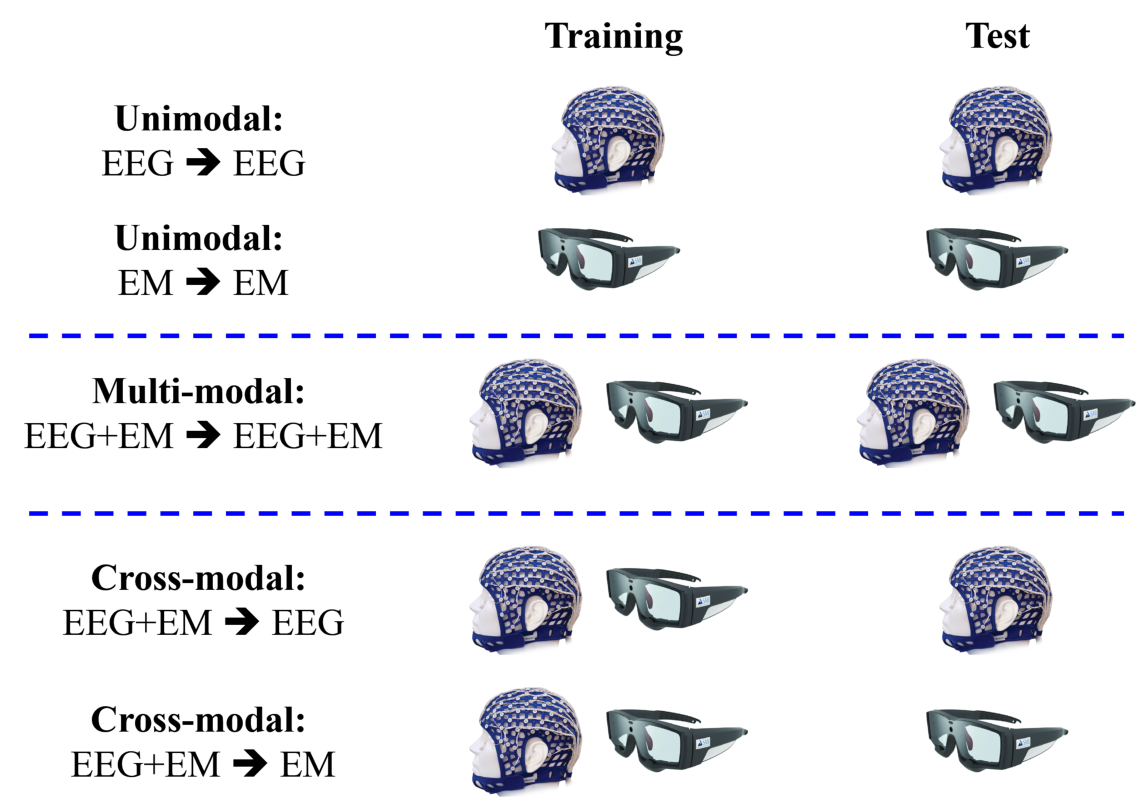}
\caption{Comparison of unimodal, multi-modal and cross-modal emotion recognition. EM stands for eye movement.} \label{fig:modal}
\end{figure}

While multi-modal fusion can improve emotion recognition accuracy, it also introduces drawbacks. Collecting multi-modal data is complex and requires multiple devices, which may be inconvenient for daily use~\cite{fei2022cross}. A multi-modal fusion model typically has more parameters, which increase the computational cost. To address these challenges and leverage the strengths of the EEG signals, we propose cross-modal supervised contrastive representation distillation (CMCRD). Fig.~\ref{fig:modal} shows a comparison of unimodal, multi-modal and cross-modal emotion recognition. In the cross-modal scenario, a pre-trained eye movement classification model assists the training of an EEG classification model, improving feature extraction from EEG signals, or vice versa. During test, only EEG signals (or eye movement signals) are acquired, eliminating the need for multi-modal data. Our approach not only improves the accuracy of emotion recognition, but also makes the system more simplified and practical. Experiments using three different neural network architectures on three multi-modal emotion recognition datasets demonstrated the effectiveness of CMCRD.

The remainder of this paper is organized as follows: Section~\ref{sect:related} presents related works. Section~\ref{sect:CMCRD} introduces our proposed CMCRD. Section~\ref{sect:exp} presents experiment results. Finally, Section~\ref{sect:conclusions} draws conclusions and points out future research directions.

\section{Related Work} \label{sect:related}

This section first introduces the commonly used multi-modal and cross-modal algorithms in emotion recognition, and then provides an overview of several knowledge distillation algorithms.

\subsection{Multi-modal Emotion Recognition}

Human emotions are a highly complex physiological process. Therefore, multiple modalities are integrated to help improve the accuracy of emotion recognition. Many studies have demonstrated that multi-modal fusion approaches for emotion recognition outperform approaches based on a single modality~\cite{zheng2014multimodal,zhu2022mutual,fu2023novel}. Recent multi-modal emotion recognition studies highlight EEG and eye movement signals as key indicators of external behavior and internal physiology, with a significant increase in research combining these signals for classification. Zheng~\emph{et al.}~\cite{zheng2014multimodal} were the first to employ a multi-modal emotion recognition framework at the feature fusion level for the recognition of three-class emotion, demonstrating that eye movement signals and EEG signals are complementary modalities, their features enhancing each other. Inspired by this pioneering work, Lu~\emph{et al.}~\cite{lu2015combining} demonstrated the inherent relationship between the two modalities and attempted to further improve emotion recognition accuracy by utilizing various modality fusion strategies. Liu~\emph{et al.}~\cite{liu2016emotion} employed a multi-modal autoencoder to extract high-dimensional interactions between modalities for feature fusion, aiming to model the interactions of multi-modal physiological signals. Liu~\emph{et al.}~\cite{liu2021comparing} used weighted summation and attention-based aggregation strategies to capture the interactions between modalities, thereby enabling more accurate emotion recognition. These studies suggest that the fusion of EEG and eye movement signals appears to be a reliable approach that can significantly improve performance.

However, multi-modal models also have shortcomings. The challenges in multi-modal data collection and the high hardware requirements have consistently limited the application of multi-modal models.

\subsection{Cross-Modal Emotion Recognition}

In recent years, to alleviate the challenges of multi-modal models, more researchers are turning to cross-modal learning, using multi-modal data only during training and single-modal data during testing. This approach facilitates model application and deployment. Spampinato~\emph{et al.}~\cite{spampinato2017deep} adopted an RNN-based approach to learn EEG data evoked by visual stimuli and used a CNN-based approach to map images to these EEG representations, allowing automatic visual classification in the brain-based visual object manifold. Jiang~\emph{et al.}~\cite{jiang2019generating} tried to use cross-modal approaches in the field of emotion recognition. Yan~\emph{et al.}~\cite{yan2021simplifying} were among the first to explore the use of conditional generative adversarial networks, attempting to generate the corresponding EEG signals from eye movement data during testing, thus eliminating the dependence on EEG signals during the testing phase. Fei~\emph{et al.}~\cite{fei2022cross} aligned different modalities in a common latent space, using correlation analysis constraints to perform cross-modal tasks, which also demonstrate superior performance. Liu~\emph{et al.}~\cite{liu2023emotionkd} developed an adaptive feedback mechanism for cross-modal knowledge distillation to enhance the classification performance of the unimodal student model by allowing the multi-modal teacher model to adjust knowledge transfer adaptively. Du~\emph{et al.}~\cite{du2023decoding} applied the Mixture-of-Products-of-Experts (MoPoE) formulation to align the latent representations of images, text, and EEG signals. This alignment enables image and text classifiers to classify latent EEG representations of images, allowing the EEG model to identify categories not used in the training from EEG signals. Fu~\emph{et al.}~\cite{fu2024cross} achieved cross-modal learning by means of feature guidance, utilizing the feature re-weighting module.

\subsection{Knowledge Distillation}

Since the introduction of the concept of knowledge distillation~\cite{hinton2015distilling}, cross-modal approaches leveraging it have been successfully validated in various fields, such as computer vision and natural language processing. Fully Information Transfer Network (FitNet) achieves knowledge transfer by passing intermediate layer features~\cite{romero2014fitnets}. Neuron Selectivity Transfer (NST) guides the student model to imitate the selective activations of neurons in the teacher model to facilitate knowledge transfer~\cite{huang2017like}. Probabilistic Knowledge Transfer (PKT) transfers knowledge by guiding the student model to learn the probability distribution of the teacher model~\cite{passalis2018learning}. Tung~\emph{et al.}~\cite{tung2019similarity} introduced Similarity-preserving (SP) knowledge distillation, which emphasizes the similarity of intermediate layer features. Relational Knowledge Distillation (RKD) focuses on relational information between the teacher and student models~\cite{park2019relational}. Tian~\emph{et al.}~\cite{tian2019contrastive} attempted to apply Contrastive Representation Distillation (CRD) to achieve cross-modal transfer from RGB images to depth images, while Arandjelovic~\emph{et al.}~\cite{arandjelovic2018objects} implemented cross-modal transfer between sound and images by AVE-Net.

However, research on utilizing knowledge distillation for cross-modal learning tasks in emotion recognition based on EEG signals remains limited. In the following sections, we will introduce the knowledge distillation approach that we employed to achieve cross-modal learning.

\section{CMCRD} \label{sect:CMCRD}

This section introduces the proposed CMCRD algorithm, which uses eye movement to guide the training of the EEG classifier. The code is available at https://github.com/kssyyy/CMCRD.

\subsection{Overview of CMCRD}

\begin{figure}[htbp]         \centering
\includegraphics[width=.8\linewidth,clip]{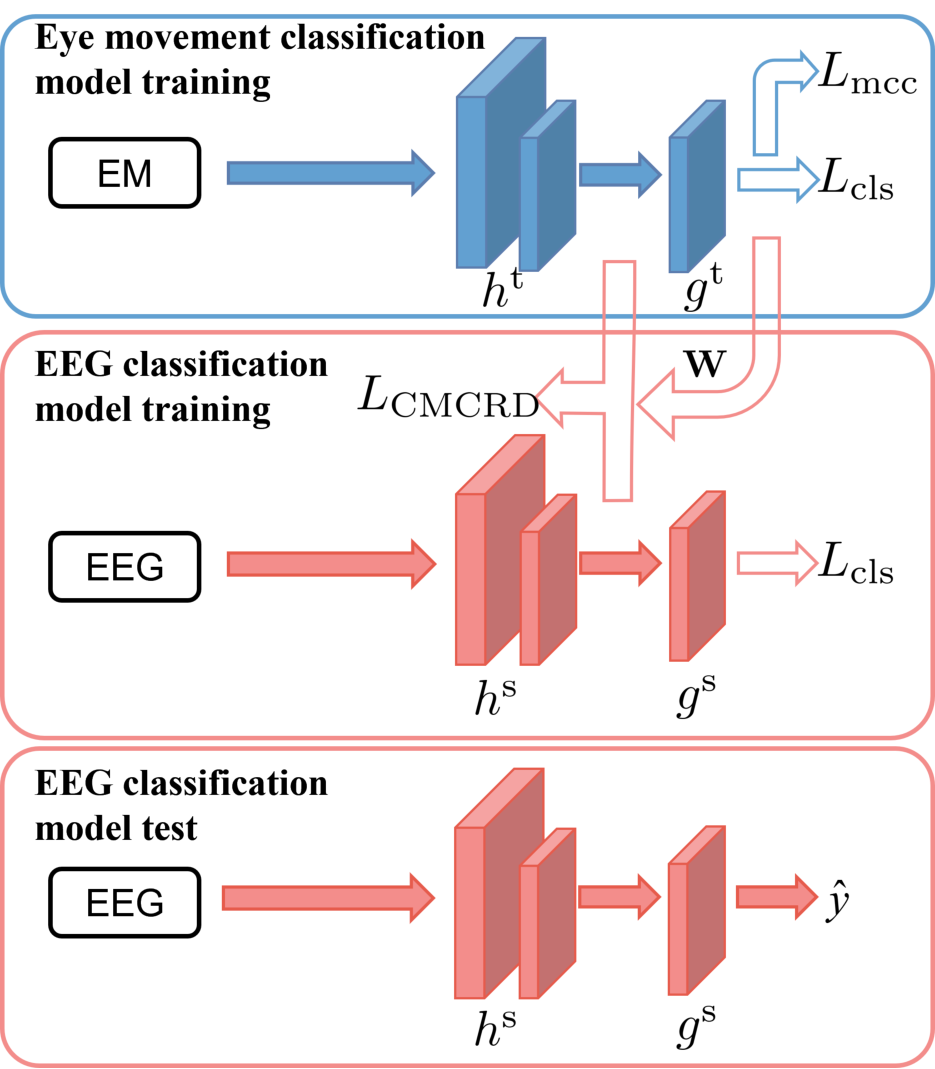}
\caption{The architecture of CMCRD. Solid arrows represent the data flow, and hollow arrows represent the gradient flow. Both an eye movement (EM) classification model and an EEG classification model are trained, but only the EEG classification model is used in testing.} \label{fig:frame}
\end{figure}

Fig.~\ref{fig:frame} shows the architecture of CMCRD, which includes an eye movement classification model and an EEG classification model. During training, both eye movement and EEG signals are available; the eye movement signal is used to train the teacher model, which then guides the training of the EEG classification model. During testing, only the EEG input is available, and hence only the EEG classification model is used.

\subsection{Teacher Model Training}

Only eye movement signals are utilized in training the teacher model, which consists of two parts: a feature extractor $h^\mathrm{t}$, and a classifier $g^\mathrm{t}$. Both parts are multilayer perceptrons. Minimum class confusion \cite{jin2020minimum} is used to reduce the distribution differences among subjects.

Let $\hat{\mathbf{y}}_{i\cdot}^\mathrm{t}=[\hat{y}_{i1}^\mathrm{t},...,\hat{y}_{ic}^\mathrm{t}]^\top$ be the logit outputs of $g^\mathrm{t}$ for an input $\mathbf{x}_i^\mathrm{t}$, where $c$ is the number of classes. The information entropy is computed as a measure of uncertainty:
\begin{align}
    H(\hat{\mathbf{y}}_{i\cdot}^\mathrm{t})=-\sum_{j=1}^{c}{\hat{y}_{ij}^\mathrm{t}\log{\hat{y}_{ij}^\mathrm{t}}}. \label{MCC_h}
\end{align}

$H(\hat{\mathbf{y}}_i^\mathrm{t})$ is then transformed into the importance of $\mathbf{x}_i^\mathrm{t}$ in computing the class confusion:
\begin{align}
    W_{ii}=\frac{n(1+\exp(-H(\hat{\mathbf{y}}_{i\cdot}^\mathrm{t}))}{\sum_{j=1}^n{(1+\exp(-H(\hat{\mathbf{y}}_{j\cdot}^\mathrm{t})))}}.\label{MCC_W}
\end{align}
Here $W_{ii}$ is the $i$-th diagonal element of the diagonal weight matrix $\mathbf{W}$, and $n$ is the batch size.

The class confusion is defined as:
\begin{align}
    C_{ij}=\hat{\mathbf{y}}_{\cdot i}^{\mathrm{t}\top} \mathbf{W}\hat{\mathbf{y}}_{\cdot j}^\mathrm{t},
\end{align}
where $\hat{\mathbf{y}}_{\cdot j}^\mathrm{t}=[\hat{y}_{1j}^\mathrm{t},...,\hat{y}_{nj}^\mathrm{t}]^\top$ contains the logits of class $j$ for all $n$ samples in the batch.

Category normalization is further applied to alleviate class-imbalance:
\begin{align}
    \Tilde{C}_{ij}=\frac{C_{ij}}{\sum_{k=1}^c C_{ik}}.
\end{align}

Finally, the minimum class confusion loss is computed as:
\begin{align}
    L_{\mathrm{mcc}}=\frac{1}{c}\sum_{i=1}^{c}\sum_{j\ne i}^{c}\lvert \Tilde{C}_{ij}\rvert.
\end{align}

The overall loss function for training the teacher model is:
\begin{align}
L^\mathrm{t}=L_{\mathrm{cls}}+\lambda_1 L_{\mathrm{mcc}},   \label{Lt}
\end{align}
where $L_{\mathrm{cls}}$ is the cross-entropy loss, and $\lambda_1$ is a trade-off parameter.

\subsection{Student Model Training}\label{subsect:Student}

CRD~\cite{tian2019contrastive} is widely employed for cross-modal representation transfer, by enhancing the lower bound of mutual information between the teacher and student model representations. However, CRD requires a large amount of data to construct contrastive learning samples, which may not be available in EEG datasets. Furthermore, it requires the student model to match the teacher's output even if it is wrong, which may cause incorrect knowledge transfer.

This subsection introduces CMCRD, an improvement of CRD, to training the student model, which consists of a feature extractor $h^\mathrm{s}$ and a classifier $g^\mathrm{s}$.

Let $\mathbf{x}_i^\mathrm{t}$ and $\mathbf{x}_i^\mathrm{s}$ be the $i$-th sample from the teacher and student modalities, respectively. Define $T=\{h^\mathrm{t}(\mathbf{x}_i^\mathrm{t})\}_{i=1}^M$ and $S=\{h^\mathrm{s}(\mathbf{x}_i^\mathrm{s})\}_{i=1}^M$, where $M$ is the total number of samples in the dataset.
Then, $k(T, S)$, the differences between the input features, can be expressed as:
\begin{align}
k(T,S)=\frac{e^{l^\mathrm{t}(T)l^\mathrm{s}(S)/\tau}}{e^{l^\mathrm{t}(T)l^\mathrm{s}(S)/\tau}+\frac{N}{M}}, \label{hts}
\end{align}
where $l^\mathrm{t}$ and $l^\mathrm{s}$ are two linear layers to map the extracted features of the teacher and the student models to the same dimensionality, $\tau$ is the temperature used to adjust the concentration level, and $N$ is the number of negative samples (those from different categories) corresponding to each sample.

Let $p(T,S)$ be the joint probability distribution, $p(T)p(S)$ be the product of marginal probability distributions, and $D$ be the indicator whether the teacher and student model inputs belong to the same class, i.e., $D = 1$ if they do, and $D = 0$ if not. Define
\begin{align}
q(T,S|D=1)&=p(T,S),\\
 q(T,S|D=0)&=p(T)p(S),
\end{align}
and
\begin{align}
\begin{split}
L(k(T,S)) & =  \mathbb{E}_{q(T,S|D=1)}\log{k(T,S)} \\
     & + N \mathbb{E}_{q(T,S|D=0)}\log{(1-k(T,S))},
\end{split}
\end{align}
i.e., $\mathbb{E}_{q(T,S|D=1)}$ is the conditional expectation under the joint probability distribution, and $\mathbb{E}_{q(T,S|D=0)}$ the conditional expectation under the product of marginal distributions.

Then, the lower bound of the mutual information between $T$ and $S$ can be expressed as~\cite{tian2019contrastive}:
\begin{align}
I(T,S)\ge \log{N}+L(k(T,S)). \label{its}
\end{align}

To increase the lower bound and hence the performance of the student model, CMCRD minimizes $-L(k(T,S))$ in its loss function. Additionally, CMCRD also introduces weighted unimodal supervised contrastive learning to maximize the utilization of limited samples.

Unlike CRD, in CMCRD's contrastive learning process, the predictions of the teacher model are constructed into a positive guidance set $\mathbb{P}$ and a negative guidance set $\mathbb{N}$:
\begin{align}
\mathbb{P} &= \left\{ (h^\mathrm{t}(\mathbf{x}_i^\mathrm{t}), h^\mathrm{s}(\mathbf{x}_i^\mathrm{s})) \mid  g^\mathrm{t}(h^\mathrm{t}(\mathbf{x}_i^\mathrm{t})) = y_i, \forall i  \right\},\\
\mathbb{N} &= \left\{ (h^\mathrm{t}(\mathbf{x}_i^\mathrm{t}), h^\mathrm{s}(\mathbf{x}_i^\mathrm{s})) \mid  g^\mathrm{t}(h^\mathrm{t}(\mathbf{x}_i^\mathrm{t})) \neq y_i, \forall i  \right\}.
\end{align}
where $y_i$ is the true class label of $\mathbf{x}_i^\mathrm{t}$.

\begin{figure}[htbp]         \centering
\includegraphics[width=.8\linewidth,clip]{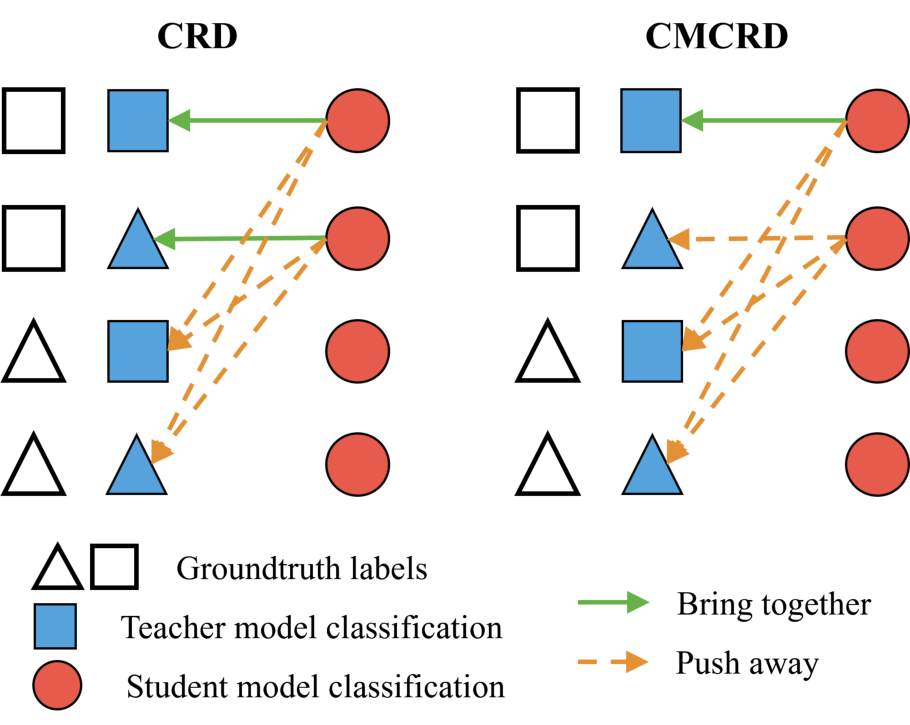}
\caption{Contrastive learning in CRD and CMCRD.} \label{fig:CMCRD}
\end{figure}

As illustrated in Fig.~\ref{fig:CMCRD}, if the teacher model's representation is classified correctly, then the student model should mimic it to improve the classification accuracy. Otherwise, the student model's representation should be different.

The CMCRD loss is then:
\small
\begin{align}
L_{\mathrm{CMCRD}} = - \log \frac{\sum\limits_{(h^\mathrm{t}(\mathbf{x}_i^\mathrm{t}), h^\mathrm{s}(\mathbf{x}_i^\mathrm{s})) \in \mathbb{P}}  \left(L(k(h^\mathrm{t}(\mathbf{x}_i^\mathrm{t}), h^\mathrm{s}(\mathbf{x}_i^\mathrm{s})))\cdot W_{ii}/ \tau\right)}
{\sum\limits_{(h^\mathrm{t}(\mathbf{x}_i^\mathrm{t}), h^\mathrm{s}(\mathbf{x}_i^\mathrm{s})) \in \mathbb{N}}  \left(L(k(h^\mathrm{t}(\mathbf{x}_i^\mathrm{t}), h^\mathrm{s}(\mathbf{x}_i^\mathrm{s})))\cdot W_{ii}/ \tau\right)},\label{lts}
\end{align}\normalsize
where $W_{ii}$ has been defined in (\ref{MCC_W}), and $\tau$ is the distillation temperature. Sampling weighting using $W_{ii}$ enables the student model to extract more knowledge from the more confident samples. As will be demonstrated in Section~\ref{sect:ablation}, sampling weighting improves the student model performance.

The overall loss function for the student model is:
\begin{align}
L^\mathrm{s}=L_{\mathrm{cls}}+\lambda_2 L_{\mathrm{CMCRD}}, \label{ls}
\end{align}
where $L_{\mathrm{cls}}$ is the cross-entropy loss, and $\lambda_2$ a trade-off parameter.

\section{Experiments} \label{sect:exp}

This section presents experiment results to validate the performance of CMCRD.

\subsection{Datasets}

Three public multi-modal datasets for emotion recognition were used:
\begin{enumerate}
\item SEED \cite{duan2013differential,zheng2015investigating}: This dataset contains 15 participants (7 males and 8 females), each with three sessions. In each session, each participants watched 15 film clips, with 5 clips from each of the three emotional categories: happy, sad, and neutral.
\item SEED-IV \cite{zheng2018emotionmeter}: This dataset also contains 15 participants (7 males and 8 females), each with three sessions. In each session, each participant watched 24 film clips, with 6 clips from each of the four emotional categories: happy, sad, fear, and neutral.
\item SEED-V \cite{liu2021comparing}: This dataset includes 20 participants (9 males and 11 females), each with three sessions. In each session, each participant watched 15 film clips, with 3 clips from each of the five emotional categories: happy, sad, fearful, disgusted, and neutral.
\end{enumerate}
Eye movement signals and 62-channel EEG signals were recorded in all three datasets.

The raw EEG signals were downsampled to 200 Hz, and then bandpass filtered between 0 and 75 Hz with baseline correction. Based on non-overlapping 4-second time windows, differential entropy (DE) features were computed across five frequency bands ($\delta: 1-3$ Hz, $\theta: 4-7$ Hz, $\alpha: 8-13$ Hz, $\beta: 14-30$ Hz, $\gamma: 31-50$ Hz). Thus, the EEG features had $62\times5=310$ dimensions.

The eye movement signals had 33/31/33 features for the three datasets, respectively, including pupil diameter, saccade dispersion, fixation dispersion, fixation duration, blink duration, saccade duration, saccade amplitude, blink frequency, and maximum fixation duration, among others.

\subsection{Experiment Setup}

Following previous studies~\cite{zhong2020eeg,li2020novel}, we conducted both within-subject and cross-subject classification experiments on all three datasets.

In within-subject classification, for each dataset, data from all three sessions were used. The first 9/16/10 trials were used for training, and the last 6/8/5 trials for testing, for the three datasets, respectively. The average classification accuracies on all subjects are reported, along with their standard deviations.

In cross-subject classification, leave-one-subject-out cross-validation was performed. In each iteration, one subject was chosen as the test set while all others were used for training and validation. The average classification accuracies across all subjects, and the standard deviations, are reported.

CMCRD was compared with several representative knowledge distillation algorithms, including CRD~\cite{tian2019contrastive}, KD~\cite{hinton2015distilling}, FitNet~\cite{romero2014fitnets}, NST~\cite{huang2017like}, SP~\cite{tung2019similarity}, RKD~\cite{park2019relational} and PKT~\cite{passalis2018learning}. Each algorithm was applied to three representative networks, DNN (6-layer MLP), DGCNN~\cite{song2018eeg}, and BiHDH~\cite{li2020novel}, which have demonstrated effectiveness in EEG emotion classification. L2 regularization was used to alleviate overfitting. All models were trained using mini-batch gradient descent, with learning rate $10^{-3}$ and batch size 32.

For CMCRD, the trade-off parameters $\lambda_1$ in (\ref{Lt}) and $\lambda_2$ in (\ref{ls}) were set to $0.1$ and $0.02$, respectively. The temperature coefficient $\tau$ in (\ref{hts}) was set to $0.07$. As the three datasets have different sizes, the number of negative samples $N$ in (\ref{its}) in within-subject experiments was set to 900, 1,200, and 900 for SEED, SEED-IV, and SEED-V, respectively, and 21,000, 25,000, and 17,000 in cross-subject experiments.

\subsection{Within-Subject Experiment Results}

Table~\ref{tab:dependent} presents the within-subject classification results on the three datasets, including those from single-modal classification (first two rows), baseline algorithms (middle section), and CMCRD (last row). Compared with EEG-only models, the classification accuracy of CMCRD increased by about 3.5\% on SEED, 3\% on SEED-IV, and 10\% on SEED-V. Compared with other knowledge distillation algorithms, CMCRD achieved the best performance across all datasets on all models. More specifically, on average it outperformed CRD by about 4\%, and the best performing baseline by about 2\%.

Interestingly, Table~\ref{tab:dependent} also shows that, while the eye movement only model did not perform well, knowledge distillation from eye movement still enhanced the EEG classification accuracy.

\begin{table*}[htbp]
\centering
\caption{Mean and standard deviations of within-subject classification accuracies (\%). EEG: EEG-only model; EM: eye movement only model.}\label{tab:dependent}
{\fontsize{9}{11}\selectfont
\resizebox{\linewidth}{!}{%
\begin{tabular}{c|ccc|ccc|ccc}
\toprule
\multirow{2}{*}{Algorithm} & & SEED&  & & SEED-IV &  &  & SEED-V & \\ \cline{2-10}
 & DNN& DGCNN& BiHDM & DNN& DGCNN & BiHDM & DNN & DGCNN & BiHDM\\
\midrule
EEG & $78.90_{\pm6.57}$ & $85.34_{\pm10.62}$ & $88.34_{\pm9.59}$ & $63.87_{\pm10.06}$ & $68.45_{\pm12.31}$ & $72.79_{\pm11.68}$ & $63.87_{\pm14.57}$ & $63.68_{\pm13.01}$ & $65.64_{\pm17.31}$\\

EM & $74.38_{\pm8.41}$ & $80.85_{\pm13.33}$ & $81.76_{\pm15.93}$ & $64.34_{\pm16.08}$ & $65.84_{\pm11.03}$ & $67.32_{\pm13.47}$ & $58.44_{\pm11.08}$ & $60.40_{\pm10.08}$ & $60.48_{\pm9.59}$\\
\midrule
CRD & $80.84_{\pm10.89}$ & $86.50_{\pm8.77}$ & $86.12_{\pm11.43}$ & $67.10_{\pm13.28}$ & $68.19_{\pm15.04}$ & $73.18_{\pm12.08}$ & $68.32_{\pm11.21}$ & $64.98_{\pm10.73}$ & $68.51_{\pm14.06}$\\

KD & $80.58_{\pm9.55}$ & $88.53_{\pm8.92}$ & $86.26_{\pm9.73}$ & $66.47_{\pm13.85}$ & $69.48_{\pm15.57}$ & $73.25_{\pm12.65}$ & $67.87_{\pm13.27}$ & $64.65_{\pm13.14}$ & $70.13_{\pm16.71}$\\

FitNet & $81.17_{\pm9.90}$ & $85.31_{\pm8.76}$ & $86.63_{\pm10.42}$ & $66.93_{\pm12.49}$ & $68.67_{\pm11.45}$ & $72.81_{\pm8.73}$ & $69.30_{\pm11.33}$ & $63.86_{\pm10.74}$ & $68.47_{\pm16.16}$\\

NST & $80.00_{\pm9.64}$ & $86.93_{\pm8.52}$ & $89.08_{\pm9.73}$ & $67.92_{\pm12.61}$ & $70.05_{\pm13.98}$ & $72.27_{\pm14.32}$ & $70.50_{\pm11.49}$ & $63.83_{\pm11.83}$ & $69.89_{\pm14.61}$\\

SP & $82.29_{\pm10.74}$ & $88.49_{\pm9.15}$ & $88.75_{\pm9.37}$ & $66.90_{\pm11.76}$ & $70.80_{\pm10.43}$ & $73.37_{\pm13.19}$ & $67.32_{\pm11.04}$ & $65.54_{\pm13.51}$ & $70.00_{\pm14.06}$\\

RKD & $81.26_{\pm8.25}$ & $87.80_{\pm7.87}$ & $89.31_{\pm9.81}$ & $66.93_{\pm14.66}$ & $70.26_{\pm11.36}$ & $73.12_{\pm10.87}$ & $66.42_{\pm13.16}$ & $65.21_{\pm12.15}$ & $70.33_{\pm17.05}$\\

PKT & $80.97_{\pm11.80}$ & $87.25_{\pm7.89}$ & $88.18_{\pm9.56}$ & $66.55_{\pm14.13}$ & $70.42_{\pm10.05}$ & $71.27_{\pm9.39}$ & $68.68_{\pm13.01}$ & $64.79_{\pm12.21}$ & $70.16_{\pm15.07}$\\
\midrule
CMCRD (ours) & $\mathbf{82.88}_{\pm9.50}$ & $\mathbf{90.27}_{\pm8.17}$ & $\mathbf{90.29}_{\pm9.93}$ & $\mathbf{68.17}_{\pm12.68}$ & $\mathbf{72.01}_{\pm14.66}$ & $\mathbf{75.25}_{\pm12.73}$ & $\mathbf{77.66}_{\pm10.63}$ & $\mathbf{74.99}_{\pm11.58}$ & $\mathbf{74.14}_{\pm11.94}$\\
\bottomrule
\end{tabular}
}}
\end{table*}

\subsection{Cross-Subject Experiment Results}

Table~\ref{tab:independent} presents the cross-subject classification accuracies on the three datasets. Compared with EEG-only models, CMCRD increased the average classification accuracy by about 3.8\% on SEED,  4.5\% on SEED-IV, and 6.8\% on SEED-V. Compared with other knowledge distillation algorithms, CMCRD achieved the best performance across almost all datasets on all models. More specifically, on average it outperformed CRD by about 2\%, and the best performing baseline by about 1.7\%.

\begin{table*}[htbp]
\centering
\caption{Mean and standard deviations of cross-subject classification accuracies (\%). EEG: EEG-only model; EM: eye movement only model.}\label{tab:independent}
{\fontsize{9}{11}\selectfont
\resizebox{\linewidth}{!}{%
\begin{tabular}{c|ccc|ccc|ccc}
\toprule
\multirow{2}{*}{Algorithm} & & SEED&  & & SEED-IV &  &  & SEED-V & \\ \cline{2-10}
 & DNN& DGCNN& BiHDM & DNN& DGCNN & BiHDM & DNN & DGCNN & BiHDM\\  \midrule
EEG & $74.12_{\pm10.01}$ & $77.49_{\pm13.92}$ & $81.51_{\pm7.00}$ & $54.09_{\pm13.42}$ & $55.48_{\pm8.67}$ & $68.33_{\pm10.22}$ & $45.86_{\pm14.68}$ & $47.69_{\pm16.23}$ & $57.14_{\pm13.33}$\\

EM & $67.87_{\pm10.74}$ & $67.32_{\pm12.69}$ & $75.45_{\pm13.10}$ & $67.58_{\pm7.62}$ & $56.44_{\pm11.40}$ & $65.43_{\pm6.26}$ & $54.06_{\pm14.27}$ & $56.43_{\pm13.51}$ & $63.72_{\pm11.66}$\\
\midrule
CRD & $76.25_{\pm8.82}$ & $78.81_{\pm12.26}$ & $83.09_{\pm10.58}$ & $58.74_{\pm14.39}$ & $57.88_{\pm9.36}$ & $69.06_{\pm8.65}$ & $52.48_{\pm19.40}$ & $53.74_{\pm14.61}$ & $58.78_{\pm15.23}$\\

KD & $76.22_{\pm9.40}$ & $77.57_{\pm13.05}$ & $81.45_{\pm9.74}$ & $57.79_{\pm13.94}$ & $55.64_{\pm12.88}$ & $69.58_{\pm9.63}$ & $49.06_{\pm12.26}$ & $48.17_{\pm14.48}$ & $59.45_{\pm12.04}$\\

FitNet & $76.28_{\pm8.90}$ & $77.66_{\pm12.01}$ & $83.04_{\pm5.98}$ & $56.90_{\pm14.21}$ & $56.71_{\pm9.72}$ & $68.58_{\pm9.71}$ & $48.33_{\pm18.10}$ & $52.61_{\pm15.95}$ & $58.37_{\pm16.12}$\\

NST & $76.77_{\pm9.45}$ & $77.28_{\pm12.60}$ & $82.80_{\pm7.54}$ & $59.40_{\pm12.69}$ & $53.20_{\pm11.84}$ & $66.86_{\pm11.43}$ & $45.44_{\pm17.11}$ & $47.97_{\pm15.83}$ & $56.29_{\pm16.41}$\\

SP & $76.59_{\pm9.76}$ & $76.72_{\pm9.75}$ & $84.73_{\pm8.08}$ & $57.24_{\pm14.77}$ & $54.70_{\pm9.89}$ & $69.35_{\pm8.66}$ & $51.42_{\pm16.05}$ & $52.93_{\pm15.73}$ & $59.19_{\pm16.85}$\\

RKD & $76.44_{\pm10.21}$ & $79.29_{\pm12.24}$ & $83.15_{\pm9.40}$ & $58.13_{\pm15.94}$ & $53.32_{\pm11.26}$ & $69.69_{\pm11.13}$ & $49.35_{\pm17.06}$ & $54.46_{\pm15.91}$ & $60.88_{\pm18.79}$\\

PKT & $76.69_{\pm8.46}$ & $77.41_{\pm14.54}$ & $82.95_{\pm8.43}$ & $56.76_{\pm11.82}$ & $55.98_{\pm9.62}$ & $69.27_{\pm11.31}$ & $48.26_{\pm16.54}$ & $53.22_{\pm15.07}$ & $60.58_{\pm16.93}$\\
\midrule
CMCRD (ours) & $\mathbf{78.38}_{\pm10.56}$ & $\mathbf{80.41}_{\pm10.41}$ & $\mathbf{85.81}_{\pm9.61}$ & $\mathbf{60.08}_{\pm14.40}$ & $\mathbf{62.04}_{\pm9.26}$ & $\mathbf{69.92}_{\pm12.57}$ & $\mathbf{52.64}_{\pm19.47}$ & $\mathbf{56.43}_{\pm17.91}$ & $\mathbf{61.87}_{\pm15.05}$\\
\bottomrule
\end{tabular}
}}
\end{table*}

We performed paired $t$-tests to evaluate whether the performance improvements of CMCRD over other knowledge distillation approaches were statistically significantly. The $p$-values, adjusted by the Benjamini-Hochberg False Discovery Rate correction, are shown in Table~\ref{tab:pvalue}. Those smaller than 0.05 are highlighted in bold. Clearly, most performance improvements of CMCRD over others were statistically significant, particularly on SEED-V. This is because SEED-V has the largest number of classes, making the classification most challenging; however, CMCRD can effectively cope with this challenge, resulting in the largest performance improvements.

\begin{table*}[htbp] \centering
\caption{Adjusted $p$-values between CMCRD and other approaches.}\label{tab:pvalue}
{\fontsize{8}{11}\selectfont
\resizebox{0.85\linewidth}{!}{%
\begin{tabular}{c|c|ccc|ccc|ccc}
\toprule
\multirow{2}{*}{Setting} & CMCRD& & SEED&  & & SEED-IV &  &  & SEED-V & \\ \cline{3-11}
 & versus & DNN& DGCNN& BiHDM & DNN& DGCNN & BiHDM & DNN & DGCNN & BiHDM\\
\midrule
\multirow{7}{*}{\begin{tabular}{c} Within- \\ Subject \end{tabular}}
&CRD    & $0.0521$ & $\mathbf{0.0235}$ & $\mathbf{0.0055}$ & $0.2571$ & $\mathbf{0.0243}$ & $0.0617$ & $\mathbf{0.0027}$ & $\mathbf{0.0001}$ & $\mathbf{0.0007}$ \\
&KD     & $\mathbf{0.0254}$ & $0.1354$ & $\mathbf{0.0060}$ & $0.0599$ & $\mathbf{0.0419}$ & $0.1610$ & $\mathbf{0.0011}$ & $\mathbf{0.0004}$ & $\mathbf{0.0461}$ \\
&FitNet & $0.1377$ & $\mathbf{0.0387}$ & $\mathbf{0.0067}$ & $0.1261$ & $\mathbf{0.0357}$ & $\mathbf{0.0268}$ & $\mathbf{0.0100}$ & $\mathbf{0.0002}$ & $\mathbf{0.0033}$ \\
&NST    & $\mathbf{0.0173}$ & $\mathbf{0.0354}$ & $0.2353$ & $0.4150$ & $0.0597$ & $\mathbf{0.0125}$ & $\mathbf{0.0175}$ & $\mathbf{0.0016}$ & $\mathbf{0.0216}$ \\
&SP     & $0.3317$ & $0.1675$ & $0.1644$ & $0.0944$ & $0.1160$ & $0.1721$ & $\mathbf{0.0033}$ & $\mathbf{0.0091}$ & $\mathbf{0.0418}$ \\
&RKD    & $0.1496$ & $0.0513$ & $0.2823$ & $0.0751$ & $0.0647$ & $0.1802$ & $\mathbf{0.0066}$ & $\mathbf{0.0012}$ & $0.0536$ \\
&PKT    & $0.0561$ & $0.0502$ & $0.0582$ & $0.0644$ & $0.0932$ & $\mathbf{0.0403}$ & $\mathbf{0.0044}$ & $\mathbf{0.0007}$ & $\mathbf{0.0338}$ \\
\midrule
\multirow{7}{*}{\begin{tabular}{c} Cross- \\ Subject \end{tabular}}
&CRD    & $\mathbf{0.0371}$ & $0.0765$ & $\mathbf{0.0391}$ & $0.3443$ & $\mathbf{0.0089}$ & $0.2172$ & $0.2062$ & $\mathbf{0.0487}$ & $\mathbf{0.0205}$ \\
&KD     & $0.0582$ & $\mathbf{0.0465}$ & $\mathbf{0.0165}$ & $0.0599$ & $\mathbf{0.0048}$ & $0.3523$ & $\mathbf{0.0352}$ & $\mathbf{0.0094}$ & $\mathbf{0.0139}$ \\
&FitNet & $0.0692$ & $\mathbf{0.0416}$ & $\mathbf{0.0328}$ & $\mathbf{0.0082}$ & $\mathbf{0.0014}$ & $0.1825$ & $\mathbf{0.0061}$ & $\mathbf{0.0355}$ & $\mathbf{0.0102}$ \\
&NST    & $0.1468$ & $\mathbf{0.0320}$ & $\mathbf{0.0493}$ & $0.6338$ & $\mathbf{0.0010}$ & $\mathbf{0.0395}$ & $\mathbf{0.0077}$ & $\mathbf{0.0121}$ & $\mathbf{0.0011}$ \\
&SP     & $0.1797$ & $\mathbf{0.0070}$ & $0.1457$ & $0.0606$ & $\mathbf{0.0048}$ & $0.3726$ & $0.1013$ & $\mathbf{0.0423}$ & $\mathbf{0.0378}$ \\
&RKD    & $0.1320$ & $0.2662$ & $0.0649$ & $0.0905$ & $\mathbf{0.0043}$ & $0.6417$ & $\mathbf{0.0398}$ & $0.1247$ & $0.2468$ \\
&PKT    & $0.1842$ & $\mathbf{0.0283}$ & $\mathbf{0.0454}$ & $\mathbf{0.0092}$ & $\mathbf{0.0017}$ & $0.2964$ & $\mathbf{0.0055}$ & $\mathbf{0.0431}$ & $0.1799$ \\
\bottomrule
\end{tabular}
}}
\end{table*}

\subsection{Feature Visualization}

\begin{figure}[htbp]         \centering
\includegraphics[width=1\linewidth,clip]{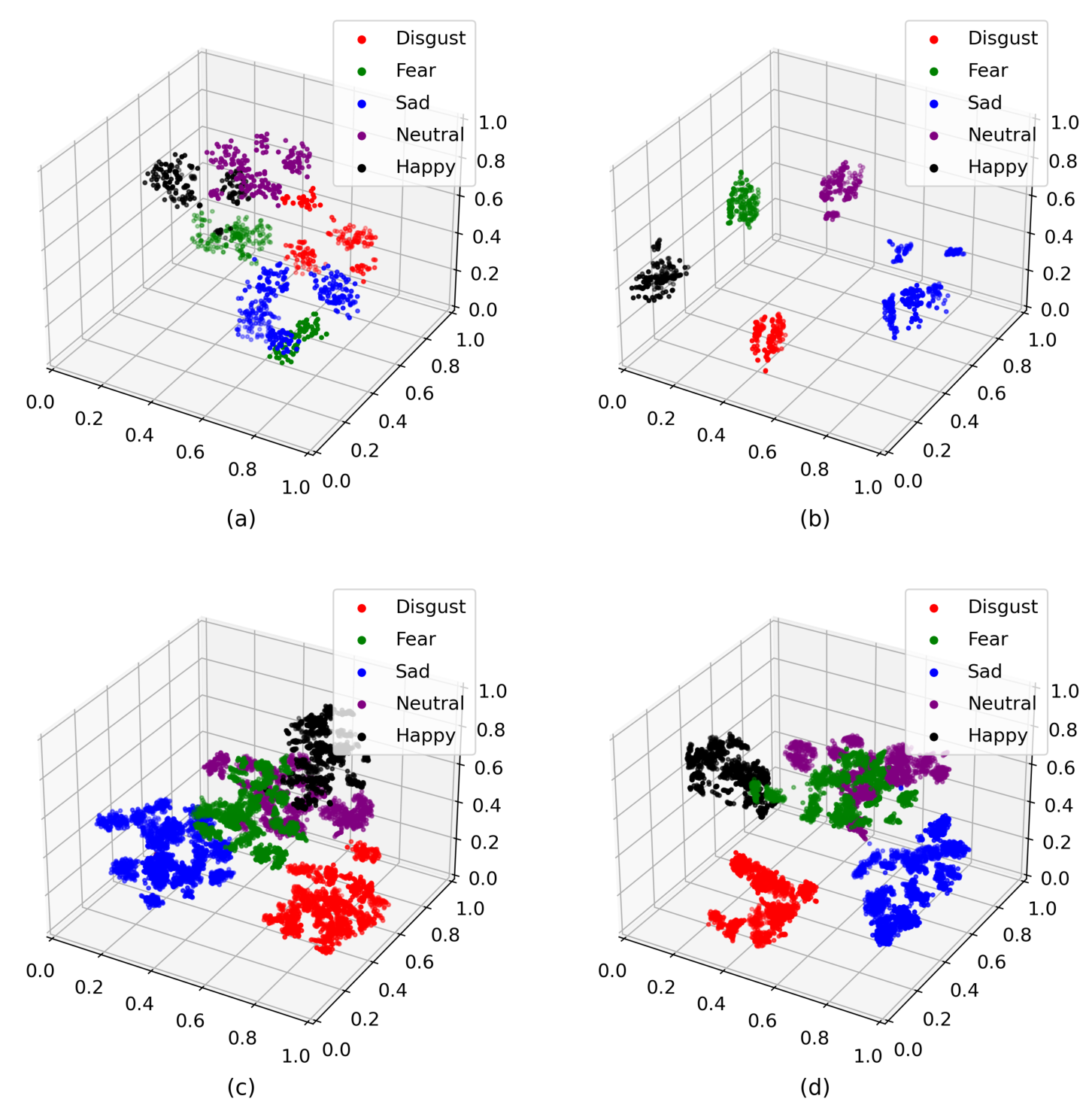}
\caption{$t$-SNE visualization of features extracted by the DNN model from the second subject in SEED-V. Different colors represent different emotions. (a) EEG-only features in within-subject setting; (b) EEG features from CMCRD in within-subject setting; (c) EEG-only features in cross-subject setting; and, (d) EEG features from CMCRD in cross-subject setting.} \label{fig:tsne}
\end{figure}

Fig.~\ref{fig:tsne} shows $t$-SNE~\cite{Maaten2008} visualization of features extracted by the DNN model from the second subject in SEED-V. The first row compares the outputs of the DNN model's feature extractor with and without the CMCRD algorithm under the subject-dependent setting, while the second row presents the same comparison under the subject-independent setting.

Comparing Fig.~\ref{fig:tsne}(a) and (b), as well as Fig.~\ref{fig:tsne}(c) and (d), it can be observed that the features extracted from CMCRD are more compact within the same cluster, and more separated among different clusters, facilitating classification.

\subsection{EEG-Guided Eye Movement Classification}

Previous experiments used eye movement signals to guide the training of EEG signals. Table~\ref{tab:compont} shows the results when EEG signals were used to guide the training of eye movement signals. CMCRD also demonstrated superior performance in this setting. Particularly, in within-subject experiments, CMCRD improved the eye movement only model by about 3.5\% on SEED, 5\% on SEED-IV, and 4\% on SEED-V. In cross-subject experiments, CMCRD improved the eye movement only models by about 2.5\% on SEED, 3\% on SEED-IV, and 5.5\% on SEED-V. On average, CMCRD outperformed CRD by about 2\%, and the best-performing knowledge distillation approach by 1.6\%.

Though the EEG only model generally outperformed the eye movement only model (since EEG has many more channels than eye movement), CMCRD achieved comparable performance with the EEG only model. This makes it very promising for real-world applications, as eye movement signals are easier to acquire than EEG signals. These results also demonstrate the versatility of CMCRD.

\begin{table*}[htbp] \centering
\caption{EEG-guided eye movement classification accuracies (\%).}\label{tab:compont}
{\fontsize{11}{13}\selectfont
\resizebox{\linewidth}{!}{%
\begin{tabular}{c|c|ccc|ccc|ccc}
\toprule
\multirow{2}{*}{Setting}& \multirow{2}{*}{Algorithm}& & SEED&  & & SEED-IV &  &  & SEED-V & \\ \cline{3-11}
 & & DNN& DGCNN& BiHDM & DNN& DGCNN & BiHDM & DNN & DGCNN & BiHDM\\ \midrule
\multirow{11}{*}{\begin{tabular}{c} Within-\\ Subject \end{tabular}} &EEG & $78.90_{\pm6.57}$ & $85.34_{\pm10.62}$ & $88.34_{\pm9.59}$ & $63.87_{\pm10.06}$ & $68.45_{\pm12.31}$ & $72.79_{\pm11.68}$ & $63.87_{\pm14.57}$ & $63.68_{\pm13.01}$ & $65.64_{\pm17.31}$\\
&EM & $74.38_{\pm8.41}$ & $80.85_{\pm13.33}$ & $81.76_{\pm15.93}$ & $64.34_{\pm16.08}$ & $65.84_{\pm11.03}$ & $67.32_{\pm13.47}$ & $58.44_{\pm11.08}$ & $60.40_{\pm10.08}$ & $60.48_{\pm9.59}$\\
\cmidrule{2-11}
&CRD & $79.00_{\pm8.00}$ & $83.96_{\pm9.26}$ & $82.43_{\pm16.88}$ & $71.58_{\pm11.08}$ & $67.24_{\pm7.45}$ & $68.21_{\pm9.82}$ & $61.10_{\pm11.16}$ & $61.59_{\pm13.34}$ & $63.05_{\pm9.41}$\\
&KD & $78.66_{\pm13.25}$ & $81.66_{\pm13.56}$ & $80.92_{\pm11.83}$ & $69.63_{\pm8.69}$ & $66.87_{\pm9.56}$ & $69.10_{\pm11.29}$ & $59.23_{\pm11.39}$ & $62.37_{\pm11.42}$ & $62.90_{\pm13.11}$\\
&FitNet & $78.60_{\pm9.03}$ & $84.05_{\pm14.45}$ & $82.19_{\pm16.02}$ & $71.22_{\pm8.02}$ & $68.65_{\pm6.47}$ & $68.56_{\pm7.19}$ & $60.70_{\pm11.53}$ & $61.76_{\pm10.66}$ & $60.14_{\pm8.46}$\\
&NST & $78.18_{\pm12.17}$ & $84.00_{\pm15.28}$ & $82.75_{\pm15.97}$ & $70.94_{\pm8.32}$ & $68.52_{\pm7.98}$ & $69.25_{\pm9.14}$ & $57.08_{\pm10.83}$ & $61.77_{\pm10.76}$ & $63.34_{\pm14.23}$\\
&SP & $78.24_{\pm10.30}$ & $83.27_{\pm12.44}$ & $82.88_{\pm15.72}$ & $67.95_{\pm9.17}$ & $66.83_{\pm9.72}$ & $69.79_{\pm7.22}$ & $59.70_{\pm11.07}$ & $62.19_{\pm11.87}$ & $64.01_{\pm9.20}$\\
&RKD & $79.05_{\pm11.47}$ & $83.73_{\pm13.00}$ & $82.89_{\pm13.68}$ & $69.76_{\pm8.31}$ & $67.08_{\pm10.46}$ & $70.17_{\pm9.83}$ & $60.54_{\pm11.28}$ & $62.10_{\pm10.95}$ & $64.56_{\pm10.53}$\\
&PKT & $77.06_{\pm10.53}$ & $83.45_{\pm12.51}$ & $82.38_{\pm15.39}$ & $70.75_{\pm7.01}$ & $67.32_{\pm7.78}$ & $69.30_{\pm8.58}$ & $59.37_{\pm11.37}$ & $62.18_{\pm11.15}$ & $63.96_{\pm10.30}$\\
\cmidrule{2-11}
&CMCRD (ours) & $\mathbf{80.63}_{\pm11.90}$ & $\mathbf{84.22}_{\pm9.99}$ & $\mathbf{83.24}_{\pm15.34}$ & $\mathbf{72.30}_{\pm6.59}$ & $\mathbf{69.13}_{\pm6.98}$ & $\mathbf{70.70}_{\pm6.57}$ & $\mathbf{61.76}_{\pm12.48}$ & $\mathbf{63.35}_{\pm12.46}$ & $\mathbf{65.16}_{\pm11.30}$\\
\midrule
\multirow{11}{*}{\begin{tabular}{c} Cross- \\ Subject \end{tabular}} &EEG & $74.12_{\pm10.01}$ & $77.49_{\pm13.92}$ & $81.51_{\pm7.00}$ & $54.09_{\pm13.42}$ & $55.48_{\pm8.67}$ & $68.33_{\pm10.22}$ & $45.86_{\pm14.68}$ & $47.69_{\pm16.23}$ & $57.14_{\pm13.33}$\\
&EM & $67.87_{\pm10.74}$ & $67.32_{\pm12.69}$ & $75.45_{\pm13.10}$ & $67.58_{\pm7.62}$ & $56.44_{\pm11.40}$ & $65.43_{\pm6.26}$ & $54.06_{\pm14.27}$ & $56.43_{\pm13.51}$ & $63.72_{\pm11.66}$\\
\cmidrule{2-11}
&CRD & $68.66_{\pm11.70}$ & $69.21_{\pm15.19}$ & $76.32_{\pm13.37}$ & $71.45_{\pm9.24}$ & $57.15_{\pm9.11}$ & $65.40_{\pm5.98}$ & $\mathbf{66.58}_{\pm11.81}$ & $57.82_{\pm9.01}$ & $67.24_{\pm13.46}$\\
&KD & $68.85_{\pm13.88}$ & $68.63_{\pm14.12}$ & $75.22_{\pm13.76}$ & $69.51_{\pm8.84}$ & $52.56_{\pm10.42}$ & $65.34_{\pm5.51}$ & $65.25_{\pm15.12}$ & $57.81_{\pm10.63}$ & $68.26_{\pm9.34}$\\
&FitNet & $68.14_{\pm13.10}$ & $69.36_{\pm10.86}$ & $76.63_{\pm10.89}$ & $71.87_{\pm9.76}$ & $59.32_{\pm9.96}$ & $64.97_{\pm4.77}$ & $64.02_{\pm13.19}$ & $57.79_{\pm11.65}$ & $68.62_{\pm11.77}$\\
&NST & $69.35_{\pm13.15}$ & $71.15_{\pm13.70}$ & $77.01_{\pm14.08}$ & $71.96_{\pm8.56}$ & $60.66_{\pm6.95}$ & $66.48_{\pm5.18}$ & $64.67_{\pm9.43}$ & $58.31_{\pm10.86}$ & $67.13_{\pm13.15}$\\
&SP & $68.17_{\pm11.68}$ & $71.17_{\pm11.21}$ & $76.76_{\pm11.82}$ & $72.14_{\pm6.98}$ & $58.23_{\pm8.32}$ & $64.90_{\pm4.62}$ & $63.64_{\pm7.84}$ & $57.94_{\pm9.25}$ & $68.24_{\pm13.07}$\\
&RKD & $69.39_{\pm13.33}$ & $71.15_{\pm11.00}$ & $76.69_{\pm15.59}$ & $70.00_{\pm7.05}$ & $61.36_{\pm6.18}$ & $66.14_{\pm5.92}$ & $63.59_{\pm13.62}$ & $58.43_{\pm8.74}$ & $68.83_{\pm13.99}$\\
&PKT & $68.05_{\pm11.77}$ & $69.34_{\pm13.51}$ & $76.03_{\pm12.08}$ & $71.87_{\pm8.92}$ & $58.83_{\pm8.47}$ & $65.78_{\pm4.53}$ & $64.47_{\pm12.14}$ & $57.48_{\pm9.34}$ & $67.15_{\pm14.67}$\\
\cmidrule{2-11}
&CMCRD (ours) & $\mathbf{69.82}_{\pm13.96}$ & $\mathbf{72.83}_{\pm15.15}$ & $\mathbf{77.87}_{\pm12.73}$ & $\mathbf{72.34}_{\pm7.05}$ & $\mathbf{64.17}_{\pm11.13}$ & $\mathbf{67.39}_{\pm4.85}$ & $66.38_{\pm9.30}$ & $\mathbf{58.79}_{\pm9.42}$ & $\mathbf{69.47}_{\pm11.88}$\\
\bottomrule
\end{tabular}
}}
\end{table*}

\subsection{Comparison with Multi-Modal Models}

We compared EEG-guided eye movement model and eye movement guided EEG model with a multi-modal model that uses decision fusion to combine the predictions of two models. Table~\ref{tab:mul} shows the results, where EEG+EM denotes the multi-modal model. Overall, the multi-modal model outperformed the eye movement guided EEG model, which in turn outperformed the EEG-guided eye movement model. However, in certain cases, the cross-modal model may outperform, or achieve comparable performance with, the multi-modal model.

\begin{table*}[htbp]
\centering
\caption{Classification accuracies (\%) of the unimodal models, multi-modal model, and cross-modal models.}\label{tab:mul}
{\fontsize{11}{13}\selectfont
\resizebox{\linewidth}{!}{%
\begin{tabular}{c|c|ccc|ccc|ccc}
\toprule
\multirow{2}{*}{Setting}& \multirow{2}{*}{Model}& & SEED&  & & SEED-IV &  &  & SEED-V & \\ \cline{3-11}
 & & DNN& DGCNN& BiHDM & DNN& DGCNN & BiHDM & DNN & DGCNN & BiHDM\\ \midrule
\multirow{5}{*}{\begin{tabular}{c} Within-\\ Subject \end{tabular}} &EEG & $78.90_{\pm6.57}$ & $85.34_{\pm10.62}$ & $88.34_{\pm9.59}$ & $63.87_{\pm10.06}$ & $68.45_{\pm12.31}$ & $72.79_{\pm11.68}$ & $63.87_{\pm14.57}$ & $63.68_{\pm13.01}$ & $65.64_{\pm17.31}$\\
&EM & $74.38_{\pm8.41}$ & $80.85_{\pm13.33}$ & $81.76_{\pm15.93}$ & $64.34_{\pm16.08}$ & $65.84_{\pm11.03}$ & $67.32_{\pm13.47}$ & $58.44_{\pm11.08}$ & $60.40_{\pm10.08}$ & $60.48_{\pm9.59}$\\
\cmidrule{2-11}
&EM + EEG & $\mathbf{86.97}_{\pm8.00}$ & $90.11_{\pm7.64}$ & $\mathbf{91.33}_{\pm8.71}$ & $\mathbf{74.65}_{\pm11.03}$ & $\mathbf{76.11}_{\pm10.42}$ & $74.18_{\pm10.49}$ & $73.64_{\pm10.36}$ & $70.71_{\pm10.79}$ & $\mathbf{77.42}_{\pm11.30}$\\
&EM-guided EEG & $82.88_{\pm9.50}$ & $\mathbf{90.27}_{\pm8.17}$ & $90.29_{\pm9.93}$ & $68.17_{\pm12.68}$ & $72.01_{\pm14.66}$ & $\mathbf{75.25}_{\pm12.73}$ & $\mathbf{77.66}_{\pm10.63}$ & $\mathbf{74.99}_{\pm11.58}$ & $74.14_{\pm11.94}$\\
&EEG-guided EM & $80.63_{\pm11.90}$ & $84.22_{\pm9.99}$ & $83.06_{\pm15.34}$ & $71.88_{\pm6.59}$ & $69.13_{\pm6.98}$ & $69.79_{\pm6.57}$ & $61.76_{\pm12.48}$ & $63.35_{\pm12.46}$ & $65.16_{\pm11.30}$\\
\midrule
\multirow{5}{*}{\begin{tabular}{c} Cross-\\ Subject \end{tabular}} &EEG & $74.12_{\pm10.01}$ & $77.49_{\pm13.92}$ & $81.51_{\pm7.00}$ & $54.09_{\pm13.42}$ & $55.48_{\pm8.67}$ & $68.33_{\pm10.22}$ & $45.86_{\pm14.68}$ & $47.69_{\pm16.23}$ & $57.14_{\pm13.33}$\\
&EM & $67.87_{\pm10.74}$ & $67.32_{\pm12.69}$ & $75.45_{\pm13.10}$ & $67.58_{\pm7.62}$ & $56.44_{\pm11.40}$ & $65.43_{\pm6.26}$ & $54.06_{\pm14.27}$ & $56.43_{\pm13.51}$ & $63.72_{\pm11.66}$\\
\cmidrule{2-11}
&EM + EEG & $\mathbf{79.17}_{\pm6.75}$ & $78.78_{\pm11.16}$ & $85.72_{\pm9.09}$ & $\mathbf{73.85}_{\pm8.51}$ & $\mathbf{69.51}_{\pm9.32}$ & $68.59_{\pm11.44}$ & $\mathbf{67.27}_{\pm12.26}$ & $\mathbf{59.25}_{\pm10.41}$ & $68.66_{\pm8.10}$\\
&EM-guided EEG& $78.38_{\pm10.56}$ & $\mathbf{80.41}_{\pm10.41}$ & $\mathbf{85.81}_{\pm9.61}$ & $60.08_{\pm14.40}$ & $62.04_{\pm9.26}$ & $\mathbf{69.92}_{\pm12.57}$ & $52.64_{\pm19.47}$ & $56.43_{\pm17.91}$ & $61.87_{\pm15.05}$\\
&EEG-guided EM & $69.82_{\pm13.96}$ & $72.83_{\pm15.15}$ & $77.73_{\pm12.73}$ & $72.34_{\pm7.05}$ & $64.17_{\pm11.13}$ & $67.39_{\pm4.85}$ & $65.82_{\pm9.30}$ & $58.79_{\pm9.42}$ & $\mathbf{69.47}_{\pm11.88}$\\
\bottomrule
\end{tabular}
}}
\end{table*}

Table~\ref{tab:time} shows the training and testing time of the DNN. The platform was a server with 251 GB RAM, an Intel Xeon Gold 8375C 2.9G Hz CPU and a NVIDIA RTX 3090 GPU. Compared with the multi-modal model, both cross-modal models required less training time, and less than half of the testing time.

In summary, the cross-modal models may sometimes achieve comparable performance with the multi-modal model, and their computational cost, particularly in testing, is much lower. Additionally, they require fewer devices to collect the physiological signals in testing. So, the cross-modal models may be preferred in real-world applications.

\begin{table}[ht]
\centering
\caption{Computation time (seconds) of different models.}\label{tab:time}
{\fontsize{10}{12}\selectfont
\resizebox{0.95\columnwidth}{!}{
\begin{tabular}{c|c|cc|cc}
\toprule
\multirow{2}{*}{Dataset}&\multirow{2}{*}{Model}&\multicolumn{2}{c|}{Within-Subject} & \multicolumn{2}{c}{Cross-Subject} \\ \cline{3-6}
&&Training  & Testing & Training & Testing \\
\midrule
\multirow{3}{*}{SEED} &EM + EEG & $30.75$ & $1.04$ & $913.28$ & $2.46$ \\
&EM-guided EEG & $\mathbf{19.18}$ & $0.46$ & $\mathbf{635.28}$ & $0.91$ \\
&EEG-guided EM & $23.46$ & $\mathbf{0.35}$ & $782.56$ & $\mathbf{0.69}$ \\
\midrule
\multirow{3}{*}{SEED-IV}&EM + EEG & $28.51$ & $0.82$ & $916.09$ & $1.85$ \\
&EM-guided EEG & $22.61$ & $0.33$ & $727.31$ & $0.97$ \\
&EEG-guided EM & $\mathbf{21.53}$ & $\mathbf{0.23}$ & $\mathbf{715.29}$ & $\mathbf{0.67}$ \\
\midrule
\multirow{3}{*}{SEED-V}&EM + EEG & $27.11$ & $0.63$ & $853.62$ & $2.00$ \\
&EM-guided EEG & $24.52$ & $0.31$ & $\mathbf{740.62}$ & $0.99$ \\
&EEG-guided EM & $\mathbf{23.49}$ & $\mathbf{0.22}$ & $758.57$ & $\mathbf{0.69}$ \\
\bottomrule
\end{tabular}
}}
\end{table}

\subsection{Ablation Study} \label{sect:ablation}

This subsection performs ablation studies on the DGCNN model to evaluate the contributions of the three modules in CMCRD, i.e., minimal class confusion in teacher model training, and unimodal supervision and information entropy weighting in student model training. Table~\ref{tab:abl} presents the results, where the three modules are referred to as MCC, US, and IEW, respectively.

\begin{table}[htpb]\centering
\caption{Ablation study results (\%). MCC: minimal class confusion in teacher model training; US: unimodal supervision in student model training; IEW: information entropy weighting in student model training.}\label{tab:abl}
{\fontsize{10}{12}\selectfont
\resizebox{\columnwidth}{!}{
\begin{tabular}{c|ccc|ccc}
\toprule
& MCC &US & IEW  & SEED & SEED-IV & SEED-V \\ \midrule
\multirow{8}{*}{\begin{tabular}{c} Within-\\ Subject \end{tabular}}&$\times$ & $\times$ & $\times$ & $85.19$ & $67.05$ & $63.30$ \\
\cmidrule{2-7}
&$\checkmark$ & $\times$ & $\times$ & $86.50$ & $68.19$ & $64.98$ \\
&$\times$ & $\checkmark$ & $\times$ & $87.09$ & $68.85$ & $66.41$ \\
&$\times$ & $\times$ & $\checkmark$ & $85.58$ & $67.35$ & $62.43$ \\
\cmidrule{2-7}
&$\checkmark$ & $\checkmark$ & $\times$ & $89.02$ & $70.87$ & $66.14$ \\
&$\checkmark$ & $\times$ & $\checkmark$ & $85.92$ & $67.95$ & $64.36$ \\
&$\times$ & $\checkmark$ & $\checkmark$ & $89.93$ & $70.48$ & $72.15$ \\
\cmidrule{2-7}
&$\checkmark$ & $\checkmark$ & $\checkmark$ & $\mathbf{90.27} $& $\mathbf{72.01} $& $\mathbf{74.99} $\\
\midrule
\multirow{8}{*}{\begin{tabular}{c} Cross-\\ Subject \end{tabular}}
&$\times$ & $\times$ & $\times$ & $76.66$ & $56.15$ & $54.03$ \\
\cmidrule{2-7}
&$\checkmark$ & $\times$ & $\times$ & $78.81$ & $57.88$ & $53.74$ \\
&$\times$ & $\checkmark$ & $\times$ & $79.26$ & $60.53$ & $55.69$ \\
&$\times$ & $\times$ & $\checkmark$ & $76.88$ & $54.77$ & $53.59$ \\
\cmidrule{2-7}
&$\checkmark$ & $\checkmark$ & $\times$ & $79.31$ & $59.13$ & $54.85$ \\
&$\checkmark$ & $\times$ & $\checkmark$ & $79.49$ & $53.98$ & $53.41$ \\
&$\times$ & $\checkmark$ & $\checkmark$ & $79.84$ & $61.16$ & $\mathbf{56.75} $\\
\cmidrule{2-7}
&$\checkmark$ & $\checkmark$ & $\checkmark$ & $\mathbf{80.41} $& $\mathbf{62.04} $& $56.43$ \\
\bottomrule
\end{tabular}
}}
\end{table}

Regardless of whether other modules are used, both minimal class confusion and unimodal supervision always improved the performance. Their combination offered even larger performance improvements. The effect of information entropy weighting itself was unstable; however, when used together with unimodal supervision, it consistently improved the performance. In summary, using all three modules together led to the best performance.

\section{Conclusions} \label{sect:conclusions}

This paper has introduced CMCRD for cross-modal emotion recognition, enhancing EEG classification accuracy by utilizing eye movement signals, or vice versa. In EEG classification model training, guidance from the eye movement signals helps extract more informative and reliable features from a limited number of labeled training samples. Additionally, we employed unimodal contrastive learning to prevent errors in the teacher model from misguiding the student model. Finally, information entropy weighting enables the student model to extract more knowledge from the more confident samples. During the testing phase, only the EEG (or eye movement) modality is required. CMCRD not only improves the emotion recognition accuracy, but also makes the system more simplified and practical. Experiments using three different neural network architectures on three multi-modal emotion recognition datasets demonstrated the effectiveness of CMCRD. Compared with the EEG-only model, it improved the average classification accuracy by about 6.2\%.

While CMCRD has demonstrated promising performance, it still has some limitations that will be considered in our future work:
\begin{enumerate}
\item  CMCRD in this paper only considers emotion classification. However, emotion recognition can also be formulated as regression problems in the 3-dimensional space of valance, arousal and dominance. It is meaningful to extend CMCRD from classification to regression.

\item CMCRD in this paper considers only the supervised training scenario. It is also meaningful to extend it to unsupervised and semi-supervised learning scenarios.

\item CMCRD does not explicitly address individual differences, which is pervasive in emotion recognition. It may be integrated with domain adaptation to cope with this challenge.
\end{enumerate}


\begin{thebibliography}{10}
\providecommand{\url}[1]{#1}
\csname url@samestyle\endcsname
\providecommand{\newblock}{\relax}
\providecommand{\bibinfo}[2]{#2}
\providecommand{\BIBentrySTDinterwordspacing}{\spaceskip=0pt\relax}
\providecommand{\BIBentryALTinterwordstretchfactor}{4}
\providecommand{\BIBentryALTinterwordspacing}{\spaceskip=\fontdimen2\font plus
\BIBentryALTinterwordstretchfactor\fontdimen3\font minus
  \fontdimen4\font\relax}
\providecommand{\BIBforeignlanguage}[2]{{%
\expandafter\ifx\csname l@#1\endcsname\relax
\typeout{** WARNING: IEEEtran.bst: No hyphenation pattern has been}%
\typeout{** loaded for the language `#1'. Using the pattern for}%
\typeout{** the default language instead.}%
\else
\language=\csname l@#1\endcsname
\fi
#2}}
\providecommand{\BIBdecl}{\relax}
\BIBdecl

\bibitem{barrett2017emotions}
L.~F. Barrett, \emph{{How} emotions are made: {The} secret life of the
  brain}.\hskip 1em plus 0.5em minus 0.4em\relax London, UK: Pan Macmillan,
  2017.

\bibitem{cowie2001emotion}
R.~Cowie, E.~Douglas-Cowie, N.~Tsapatsoulis, G.~Votsis, S.~Kollias, W.~Fellenz,
  and J.~G. Taylor, ``Emotion recognition in human-computer interaction,''
  \emph{IEEE Signal Processing Magazine}, vol.~18, no.~1, pp. 32--80, 2001.

\bibitem{lin2010eeg}
Y.-P. Lin, C.-H. Wang, T.-P. Jung, T.-L. Wu, S.-K. Jeng, J.-R. Duann, and J.-H.
  Chen, ``{EEG}-based emotion recognition in music listening,'' \emph{IEEE
  Trans. on Biomedical Engineering}, vol.~57, no.~7, pp. 1798--1806, 2010.

\bibitem{zheng2018emotionmeter}
W.-L. Zheng, W.~Liu, Y.~Lu, B.-L. Lu, and A.~Cichocki, ``Emotionmeter: {A}
  multimodal framework for recognizing human emotions,'' \emph{IEEE Trans. on
  Cybernetics}, vol.~49, no.~3, pp. 1110--1122, 2018.

\bibitem{drwuCTIAL2024}
Y.~Xu, X.~Jiang, and D.~Wu, ``Cross-task inconsistency based active learning
  ({CTIAL}) for emotion recognition,'' \emph{IEEE Trans. on Affective
  Computing}, vol.~15, no.~3, pp. 1659--1668, 2024.

\bibitem{zhong2014learning}
L.~Zhong, Q.~Liu, P.~Yang, J.~Huang, and D.~N. Metaxas, ``Learning multiscale
  active facial patches for expression analysis,'' \emph{IEEE Trans. on
  Cybernetics}, vol.~45, no.~8, pp. 1499--1510, 2014.

\bibitem{lee2019context}
J.~Lee, S.~Kim, S.~Kim, J.~Park, and K.~Sohn, ``Context-aware emotion
  recognition networks,'' in \emph{Proc. IEEE/CVF Int’l Conf. on Computer
  Vision}, Long Beach, CA, Jun. 2019, pp. 10\,143--10\,152.

\bibitem{el2011survey}
M.~El~Ayadi, M.~S. Kamel, and F.~Karray, ``Survey on speech emotion
  recognition: {Features}, classification schemes, and databases,''
  \emph{Pattern Recognition}, vol.~44, no.~3, pp. 572--587, 2011.

\bibitem{drwuTAC2022}
Y.~Xu, Y.~Cui, X.~Jiang, Y.~Yin, J.~Ding, L.~Li, and D.~Wu,
  ``Inconsistency-based multi-task cooperative learning for emotion
  recognition,'' \emph{IEEE Trans. on Affective Computing}, vol.~13, no.~4, pp.
  2017--2027, 2022.

\bibitem{soleymani2011multimodal}
M.~Soleymani, M.~Pantic, and T.~Pun, ``Multimodal emotion recognition in
  response to videos,'' \emph{IEEE Trans. on Affective Computing}, vol.~3,
  no.~2, pp. 211--223, 2011.

\bibitem{zheng2017identifying}
W.-L. Zheng, J.-Y. Zhu, and B.-L. Lu, ``Identifying stable patterns over time
  for emotion recognition from {EEG},'' \emph{IEEE Trans. on Affective
  Computing}, vol.~10, no.~3, pp. 417--429, 2017.

\bibitem{alarcao2017emotions}
S.~M. Alarcao and M.~J. Fonseca, ``Emotions recognition using {EEG} signals:
  {A} survey,'' \emph{IEEE Trans. on Affective Computing}, vol.~10, no.~3, pp.
  374--393, 2017.

\bibitem{wu2023affective}
D.~Wu, B.-L. Lu, B.~Hu, and Z.~Zeng, ``Affective {Brain--Computer Interfaces
  (aBCIs): A Tutorial},'' \emph{Proc. of the IEEE}, vol. 111, no.~10, pp.
  1314--1332, 2023.

\bibitem{wang2014emotional}
X.-W. Wang, D.~Nie, and B.-L. Lu, ``Emotional state classification from {EEG}
  data using machine learning approach,'' \emph{Neurocomputing}, vol. 129, pp.
  94--106, 2014.

\bibitem{zhang2018cascade}
D.~Zhang, L.~Yao, X.~Zhang, S.~Wang, W.~Chen, R.~Boots, and B.~Benatallah,
  ``Cascade and parallel convolutional recurrent neural networks on {EEG}-based
  intention recognition for brain computer interface,'' in \emph{Proc. AAAI
  Conf. on Artificial Ntelligence}, vol.~32, no.~1, New Orleans, LA, Feb. 2018,
  pp. 1703--1710.

\bibitem{lawhern2018eegnet}
V.~J. Lawhern, A.~J. Solon, N.~R. Waytowich, S.~M. Gordon, C.~P. Hung, and
  B.~J. Lance, ``{EEGNet}: {A} compact convolutional neural network for
  {EEG}-based brain--computer interfaces,'' \emph{Journal of Neural
  Engineering}, vol.~15, no.~5, p. 056013, 2018.

\bibitem{zhang2018spatial}
T.~Zhang, W.~Zheng, Z.~Cui, Y.~Zong, and Y.~Li, ``Spatial--temporal recurrent
  neural network for emotion recognition,'' \emph{IEEE Trans. on Cybernetics},
  vol.~49, no.~3, pp. 839--847, 2018.

\bibitem{li2020novel}
Y.~Li, L.~Wang, W.~Zheng, Y.~Zong, L.~Qi, Z.~Cui, T.~Zhang, and T.~Song, ``A
  novel bi-hemispheric discrepancy model for {EEG} emotion recognition,''
  \emph{IEEE Trans. on Cognitive and Developmental Systems}, vol.~13, no.~2,
  pp. 354--367, 2020.

\bibitem{chowdary2022emotion}
M.~K. Chowdary, J.~Anitha, and D.~J. Hemanth, ``Emotion recognition from {EEG}
  signals using recurrent neural networks,'' \emph{Electronics}, vol.~11,
  no.~15, p. 2387, 2022.

\bibitem{song2018eeg}
T.~Song, W.~Zheng, P.~Song, and Z.~Cui, ``{EEG} emotion recognition using
  dynamical graph convolutional neural networks,'' \emph{IEEE Trans. on
  Affective Computing}, vol.~11, no.~3, pp. 532--541, 2018.

\bibitem{zhong2020eeg}
P.~Zhong, D.~Wang, and C.~Miao, ``{EEG}-based emotion recognition using
  regularized graph neural networks,'' \emph{IEEE Trans. on Affective
  Computing}, vol.~13, no.~3, pp. 1290--1301, 2020.

\bibitem{song2022eeg}
Y.~Song, Q.~Zheng, B.~Liu, and X.~Gao, ``{EEG} conformer: {Convolutional}
  transformer for {EEG} decoding and visualization,'' \emph{IEEE Trans. on
  Neural Systems and Rehabilitation Engineering}, vol.~31, pp. 710--719, 2022.

\bibitem{ding2024eeg}
Y.~Ding, Y.~Li, H.~Sun, R.~Liu, C.~Tong, and C.~Guan, ``{EEG-Deformer}: {A}
  dense convolutional transformer for brain-computer interfaces,'' \emph{IEEE
  Journal of Biomedical and Health Informatics}, 2024.

\bibitem{zhu2022mutual}
J.~Zhu, C.~Yang, X.~Xie, S.~Wei, Y.~Li, X.~Li, and B.~Hu, ``Mutual information
  based fusion model {(MIBFM)}: {Mild} depression recognition using {EEG} and
  pupil area signals,'' \emph{IEEE Trans. on Affective Computing}, vol.~14,
  no.~3, pp. 2102--2115, 2022.

\bibitem{fu2023novel}
B.~Fu, C.~Gu, M.~Fu, Y.~Xia, and Y.~Liu, ``A novel feature fusion network for
  multimodal emotion recognition from {EEG} and eye movement signals,''
  \emph{Frontiers in Neuroscience}, vol.~17, p. 1234162, 2023.

\bibitem{lu2015combining}
Y.~Lu, W.-L. Zheng, B.~Li, and B.-L. Lu, ``Combining eye movements and {EEG} to
  enhance emotion recognition,'' in \emph{Proc. 24th Int’l Joint Conf. on
  Artificial Intelligence}, vol.~15, Buenos Aires, Argentina, Jul. 2015, pp.
  1170--1176.

\bibitem{zhu2022content}
J.~Zhu, S.~Wei, X.~Xie, C.~Yang, Y.~Li, X.~Li, and B.~Hu, ``Content-based
  multiple evidence fusion on {EEG} and eye movements for mild depression
  recognition,'' \emph{Computer Methods and Programs in Biomedicine}, vol. 226,
  p. 107100, 2022.

\bibitem{fei2022cross}
C.~Fei, R.~Li, L.-M. Zhao, Z.~Li, and B.-L. Lu, ``A cross-modality deep
  learning method for measuring decision confidence from eye movement
  signals,'' in \emph{Proc. 2022 44th Annual Int'l Conf. on IEEE Engineering in
  Medicine and Biology Society}, Glasgow, UK, Jul. 2022, pp. 3342--3345.

\bibitem{zheng2014multimodal}
W.-L. Zheng, B.-N. Dong, and B.-L. Lu, ``Multimodal emotion recognition using
  {EEG} and eye tracking data,'' in \emph{2014 6th Annual Int'l Conf. on IEEE
  Engineering in Medicine and Biology Society}, Chicago, IL, Aug. 2014, pp.
  5040--5043.

\bibitem{liu2016emotion}
W.~Liu, W.-L. Zheng, and B.-L. Lu, ``Emotion recognition using multimodal deep
  learning,'' in \emph{23rd Int'l Conf. on Neural Information
  Processing}.\hskip 1em plus 0.5em minus 0.4em\relax Kyoto, Japan: Springer,
  Oct. 2016, pp. 521--529.

\bibitem{liu2021comparing}
W.~Liu, J.-L. Qiu, W.-L. Zheng, and B.-L. Lu, ``Comparing recognition
  performance and robustness of multimodal deep learning models for multimodal
  emotion recognition,'' \emph{IEEE Trans. on Cognitive and Developmental
  Systems}, vol.~14, no.~2, pp. 715--729, 2021.

\bibitem{spampinato2017deep}
C.~Spampinato, S.~Palazzo, I.~Kavasidis, D.~Giordano, N.~Souly, and M.~Shah,
  ``Deep learning human mind for automated visual classification,'' in
  \emph{Proc. IEEE Conf. on Computer Vision and Pattern Recognition}, Honolulu,
  HI, Jul. 2017, pp. 6809--6817.

\bibitem{jiang2019generating}
H.~Jiang, X.~Guan, W.-Y. Zhao, L.-M. Zhao, and B.-L. Lu, ``Generating
  multimodal features for emotion classification from eye movement signals.''
  \emph{Australian Journal of Intelligent Information Processing Systems},
  vol.~15, no.~3, pp. 59--66, 2019.

\bibitem{yan2021simplifying}
X.~Yan, L.-M. Zhao, and B.-L. Lu, ``Simplifying multimodal emotion recognition
  with single eye movement modality,'' in \emph{Proc. 29th {ACM} Int'l Conf. on
  Multimedia}, Virtual Event, Oct. 2021, pp. 1057--1063.

\bibitem{liu2023emotionkd}
Y.~Liu, Z.~Jia, and H.~Wang, ``Emotion{KD}: {A} cross-modal knowledge
  distillation framework for emotion recognition based on physiological
  signals,'' in \emph{Proc. 31st ACM Int'l Conf. on Multimedia}, Ottawa,
  Canada, Oct. 2023, pp. 6122--6131.

\bibitem{du2023decoding}
C.~Du, K.~Fu, J.~Li, and H.~He, ``Decoding visual neural representations by
  multimodal learning of brain-visual-linguistic features,'' \emph{IEEE Trans.
  on Pattern Analysis and Machine Intelligence}, vol.~45, no.~9, pp.
  10\,760--10\,777, 2023.

\bibitem{fu2024cross}
B.~Fu, W.~Chu, C.~Gu, and Y.~Liu, ``Cross-modal guiding neural network for
  multimodal emotion recognition from {EEG} and eye movement signals,''
  \emph{IEEE Journal of Biomedical and Health Informatics}, vol.~28, no.~10,
  pp. 5865--5876, 2024.

\bibitem{hinton2015distilling}
G.~Hinton, ``Distilling the knowledge in a neural network,'' in \emph{Proc.
  NIPS Deep Learning and Representation Learning Workshop}, Montreal, Canada,
  Dec. 2015.

\bibitem{romero2014fitnets}
A.~Romero, N.~Ballas, S.~E. Kahou, A.~Chassang, C.~Gatta, and Y.~Bengio,
  ``Fitnets: {Hints} for thin deep nets,'' in \emph{Proc. Neural Information
  Processing Systems}, Montreal, Canada, Dec. 2014, pp. 1--9.

\bibitem{huang2017like}
Z.~Huang and N.~Wang, ``Like what you like: {Knowledge} distill via neuron
  selectivity transfer,'' in \emph{Proc. Int'l Conf. on Learning
  Representations}, New Orleans, LA, May 2019.

\bibitem{passalis2018learning}
N.~Passalis and A.~Tefas, ``Learning deep representations with probabilistic
  knowledge transfer,'' in \emph{Proc. European Conf. on Computer Vision},
  Munich, Germany, Sep. 2018, pp. 268--284.

\bibitem{tung2019similarity}
F.~Tung and G.~Mori, ``Similarity-preserving knowledge distillation,'' in
  \emph{Proc. IEEE/CVF Int'l Conf. on Computer Vision}, Seoul, South Korea,
  Oct. 2019, pp. 1365--1374.

\bibitem{park2019relational}
W.~Park, D.~Kim, Y.~Lu, and M.~Cho, ``Relational knowledge distillation,'' in
  \emph{Proc. IEEE/CVF Conf. on Computer Vision and Pattern Recognition}, Long
  Beach, CA, Jun. 2019, pp. 3967--3976.

\bibitem{tian2019contrastive}
Y.~Tian, D.~Krishnan, and P.~Isola, ``Contrastive representation
  distillation,'' in \emph{Proc. Int'l Conf. on Learning Representations}, New
  Orleans, LA, Apr. 2019.

\bibitem{arandjelovic2018objects}
R.~Arandjelovic and A.~Zisserman, ``Objects that sound,'' in \emph{Proc. Eur.
  Conf. on Computer Vision}, Munich, Germany, Sep. 2018, pp. 435--451.

\bibitem{jin2020minimum}
Y.~Jin, X.~Wang, M.~Long, and J.~Wang, ``Minimum class confusion for versatile
  domain adaptation,'' in \emph{Proc. European Conf. on Computer Vision},
  Glasgow, UK, Aug. 2020, pp. 464--480.

\bibitem{duan2013differential}
R.-N. Duan, J.-Y. Zhu, and B.-L. Lu, ``Differential entropy feature for
  {EEG}-based emotion classification,'' in \emph{2013 6th IEEE/EMBS Int'l Conf.
  on Neural Engineering}, San Diego, CA, May. 2013, pp. 81--84.

\bibitem{zheng2015investigating}
W.-L. Zheng and B.-L. Lu, ``Investigating critical frequency bands and channels
  for {EEG}-based emotion recognition with deep neural networks,'' \emph{IEEE
  Trans. on Autonomous Mental Development}, vol.~7, no.~3, pp. 162--175, 2015.

\bibitem{Maaten2008}
L.~van~der Maaten and G.~Hinton, ``Visualizing data using t-{SNE},''
  \emph{Journal of Machine Learning Research}, vol.~9, pp. 2579--2605, 2008.

\end{thebibliography}


\end{document}